\documentclass[a4paper, twocolumn]{article}
\usepackage{amsmath}
\usepackage{graphicx}
\usepackage{epstopdf}
\usepackage{multirow}
\usepackage{threeparttable}
\usepackage{caption}
\usepackage{subcaption}

\usepackage{amsmath,amsfonts,amsthm,amssymb}

\usepackage{geometry}
\usepackage{fancyhdr}

\fancypagestyle{firststyle}{
  \fancyhf{}
  \fancyhead[C]{}
}

\begin{document}

\title{Sound Field Estimation around a Rigid Sphere with Physics-informed Neural Network}
\author{Xingyu Chen, Fei Ma, Amy Bastine, Prasanga Samarasinghe, Huiyuan Sun\\Audio and Acoustic Signal Processing Group, The Australian National University, Canberra, Australia\\School of Electrical and Information Engineering, The University of Sydney, Australia} 



\maketitle
\thispagestyle{firststyle}
\pagestyle{empty}

\begin{abstract}
Accurate estimation of the sound field around a rigid sphere necessitates adequate sampling on the sphere, which may not always be possible. To overcome this challenge, this paper proposes a method for sound field estimation based on a physics-informed neural network. This approach integrates physical knowledge into the architecture and training process of the network. In contrast to other learning-based methods, the proposed method incorporates additional constraints derived from the Helmholtz equation and the zero radial velocity condition on the rigid sphere. Consequently, it can generate physically feasible estimations without requiring a large dataset. In contrast to the spherical harmonic-based method, the proposed approach has better fitting abilities and circumvents the ill condition caused by truncation. Simulation results demonstrate the effectiveness of the proposed method in achieving accurate sound field estimations from limited measurements, outperforming the spherical harmonic method and plane-wave decomposition method.
\end{abstract}

\section{Introduction}
Sound field analysis is a critical area of research with a wide range of applications, including spatial audio processing \cite{breebaart2007spatial, rumsey2012spatial}, augmented reality/virtual reality \cite{poletti2005three}, and active noise control \cite{zhang2018active, Fei_2020}. 
To determine the distribution of acoustic pressure in a given environment, researchers often use a microphone array. 
However, microphone arrays cause reflections and scattering, which potentially impacts the accuracy of the sound field estimation. 
Therefore, understanding the scattering field is essential to ensure the accuracy of the sound field estimation.

Rigid spherical microphone arrays are extensively employed in sound field analysis due to their theoretically derived scattering fields, enabling the estimation of sound fields in three-dimensional space surrounding the array \cite{williams2000fourier,rafaely2004analysis,williams2010vector}.
Conventional methods, such as the spherical harmonic (SH) method \cite{abhayapala2002theory}, represent the sound field by a set of orthonormal modes.  
However, these methods have inherent limitations in practical applications. For instance, the limited number of microphones leads to a discrete representation through integration, resulting in truncation errors \cite{ward2001reproduction}. Plane-wave decomposition method assumes that the sound field consists only of plane waves \cite{rafaely2004plane}. It neglects the spatial variation within each wavefront, which may not fully capture the intricacies of the scattering fields.

Learning-based methods have been employed in scattering problems, and have demonstrated effectiveness in fitting nonlinear mappings \cite{wei2018deep,
chen2020review}. Among these methods, Convolutional Neural Networks (CNNs) have gained popularity due to their ability to promote parameter sharing and reduce the number of trainable parameters \cite{lecun2015deep}. However, in the context of sound field estimation, convolutional and max-pooling layers introduce pixel spaces that can lead to discontinuities in sound pressure continuity. This discontinuity potentially undermines the physical interpretation of the data \cite{lluis2020sound, kristoffersen2021deep, shigemi2022physics}. Additionally, the heavy reliance on training data often necessitates a large simulated dataset. 

Recently, researchers introduced Physics-Informed Neural Networks (PINNs) as a method to incorporate physical knowledge into the network architecture and training \cite{raissi2017physics, karniadakis2021physics, shigemi2022physics, liu2023sd}. 
This integration allows PINNs to achieve physically feasible estimations and require less training data. Fully connected Neural Networks (FNNs) are commonly employed as part of PINN due to their ability to approximate any continuous function, as guaranteed by the Universal Approximation Theorem \cite{scarselli1998universal}.

In this paper, we propose a method for sound field estimation using PINNs around a rigid sphere. We employ a small FNN network structure and incorporate two constraints: the Helmholtz equation and the fact that sound pressure cannot generate radial motion at the sphere boundary. 
We use the FNN to model the mapping between Euclidean space positions and sound pressure and use these constraints as a loss function to train the network. Notably, our method relies solely on measured data from microphones, eliminating the need for setting up a grid and using simulated datasets. Simulation results demonstrate the effectiveness of our method, outperforming the spherical harmonic method and the plane-wave decomposition method.

\section{problem formulation}
We introduce two coordinate systems, as shown in Fig. 1, where $\boldsymbol{x} = (x,y,z)$ and 
$\boldsymbol{r}=(r,\theta,\phi)$ represent Cartesian and spherical coordinates, respectively. 
Our study focuses on a sound field $\mathcal{V} \subset \mathbb{R}^3$, which includes a rigid 
sphere located at the origin $O$ with a radius of $a$. 
We aim to estimate the sound field $P(\boldsymbol{r},\omega)$ at position $\boldsymbol{r}$ and angular frequency $\omega$ around the sphere $r \geq a $, by using measurements from $Q$ microphones mounted on the rigid sphere.

\begin{figure}[t]
\begin{center}
\includegraphics[width=5.5cm]{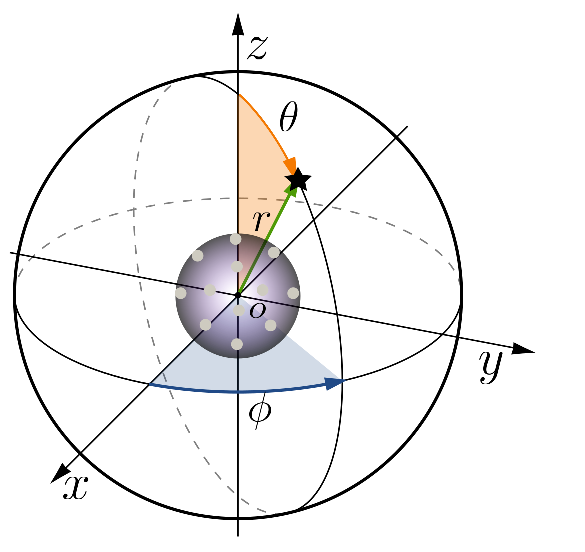}
\end{center}
\caption{System setup: A rigid sphere with microphones showing Cartesian and spherical coordinates. $\bullet$ represent the microphones, and $\star$ represents a reference point.}
\end{figure}

For a homogeneous sound field (absence of sound sources) around a rigid sphere, theoretical solutions for the sound field have been established \cite{williams2000fourier}. The spherical harmonic (SH) method decomposes $P(\boldsymbol{r},\omega)$ into a series of orthonormal modes \cite{williams2000fourier}
\begin{equation}
P(\boldsymbol{r}, \omega)=\sum_{n=0}^{\infty} G_n(r, a, \omega) \sum_{m=-n}^n \mathcal{P}_n^m(a, \omega) Y_n^m(\theta, \phi),
\label{eq:1}
\end{equation}
where $G_n(r, a, \omega)$ is the radial propagator, $\mathcal{P}_n^m(a, \omega)$ are the spherical harmonic coefficients, and $Y_n^m(\theta, \phi)$ is the spherical harmonic of order $n$ and degree $m$. 
$\mathcal{P}_n^m(a, \omega)$ are obtained by integrating $P(\mathbf{a}, \omega)$ over the surface of the sphere
\begin{equation}
\label{eq:2}
\mathcal{P}_n^m(a, \omega)=\iint P(\mathbf{a}, \omega) Y_n^m(\theta, \phi)^{*} d \Omega,
\end{equation}
where $\mathbf{a}=(a, \theta, \phi)$, 
$\Omega=(\theta,\phi)$, 
and $\iint (\cdot) d \Omega \equiv \int_{0}^{2\pi}\int_{0}^{\pi} (\cdot) \sin\theta d \theta d \phi$.
$G_n(r, a, \omega)$ must satisfy the boundary condition on the rigid sphere, leading to 

\begin{equation}
G_n(r, a, \omega)=j_n(\frac{\omega r}{c})-\frac{j_n^{\prime}\left(\frac{\omega a}{c}\right)}{h_n^{(2)^{\prime}}\left(\frac{\omega a}{c}\right)} h_n^{(2)}(\frac{\omega r}{c}),
\label{eq:3}
\end{equation}
where $j_n(\cdot)$ and $h_n^{(2)}(\cdot)$ are the spherical Bessel function of the first kind and the spherical Hankel function of the second kind, respectively, and $j_n^{\prime}(\cdot)$ and $h_n^{(2)^{\prime}}(\cdot)$ are corresponding derivatives \cite{williams2010vector}, $c$ is the speed of sound.

There are three problems with the implementation of the theoretical method.
First, the infinite series in (\ref{eq:1}) needs to be truncated at limited order $N$ \cite{ward2001reproduction}. 
Second, the pressure on the sphere is sampled at a limited ($Q$) number of positions. 
Hence, the integral in (\ref{eq:2}) needs to be discretely approximated. 
Third, in the low-frequency region where the distance $r$ increases ($r > a$), the propagators $G_n(r, a, \omega)$ approximately follow a power law of the form $\frac{n+1}{2n+1} \left(\frac{r}{a}\right)^n$. These increasing values have significant implications as they indicate an ill-posed problem \cite{williams2010vector}. Consequently, small variations in $\mathcal{P}_n^m(a,\omega)$ resulting from truncation errors or noise will be magnified through the propagator.

To overcome these challenges and obtain more accurate estimations, we propose a neural network approach integrated with physics laws. This method aims to establish a regression relationship between coordinates and sound pressure while satisfying the governing physics equations.

\section{PINN method}
In this section, we present the PINN method, which consists of three steps: (1) constructing an FNN using microphone positions and measured sound pressure as training pairs, (2) incorporating physical knowledge as a constraint in the network, and (3) adjusting the weight of the loss function to facilitate convergence. The workflow of the PINN method is presented in Fig. 2.

FNN is a composition of multiple layers, and the nodes between adjacent layers are fully connected.
A single layer of the network is denoted as 
\begin{equation}
F(\boldsymbol{x})=\sigma\left(\boldsymbol{x}^{\intercal} \mathbf{w}+b\right),
\end{equation}
where $\boldsymbol{x} \in \mathbb{R}^3$ is the inputs, which in this case are the Cartesian positions in the sound field $\mathcal{V}$, $\mathbf{w}$ represents the weights, $b$ is the bias, and $\sigma$ is the activation function. The FNN is a composition of multiple layers
\begin{equation}
\Phi(\boldsymbol{x}, \omega ;  \psi)=\left(F_n \circ \ldots \circ F_2 \circ F_1\right)(\boldsymbol{x}),
\end{equation}
where $\psi$ represents the set of all trainable parameters $\mathbf{w}$ and $b$. 
Hereafter, $\omega$ is omitted in $\Phi(\boldsymbol{x};\psi)$ when representing acoustic quantities for notional simplicity.
$ \psi$ is learned by minimizing the mean squared error (MSE) loss $\mathcal{L}_{\mathrm{data}}$ with respect to the training data
\begin{equation}
\psi^*=\underset{ \psi}{\arg \min} \; \mathcal{L}_{\mathrm{data}}(\boldsymbol{x}; \psi),
\end{equation}
where $\mathcal{L}_{\mathrm{data}}$ is defined as
\begin{equation}
\mathcal{L}_{\mathrm{data}}(\boldsymbol{x}_i; \psi)=\frac{1}{Q} \sum_{i=1}^Q\left(P(\boldsymbol{x}_i,\omega)-\Phi\left(\boldsymbol{x}_i;  \psi\right)\right)^2 .
\end{equation} 
It fits the regression relationship between the spatial coordinates of the microphones $\boldsymbol{x}$ and the corresponding sound pressure values $P(\boldsymbol{x},\omega)$. The network results $\Phi(\boldsymbol{x};\psi)$ represent the estimated sound pressure $\hat{P}(\boldsymbol{x},\omega)$.

The aforementioned approach represents a purely data-driven method. 
It needs a large dataset, which is not always possible.
We incorporate domain knowledge to reduce the data requirements and generate physically meaningful estimations.

For sound field estimation around a rigid sphere, we can leverage two domain knowledge to improve the estimation accuracy. 
Firstly, sound fields are governed by the Helmholtz equation
\begin{equation}
\left(\triangle+(\frac{c}{\omega})^2\right) P(\boldsymbol{r},\omega)=0,
\label{eq:7}
\end{equation}
where $\triangle = \partial^2 / \partial x^2+\partial^2 / \partial y^2+\partial^2 / \partial z^2$ 
denotes the Laplace operator. 
Therefore, we require the PINN estimation to satisfy \eqref{eq:7}, i.e.,
\begin{equation}
\left(\triangle+(\frac{c}{\omega})^2\right) \Phi(\boldsymbol{x};  \psi)=0.
\label{eq:8}
\end{equation}
Second, radial velocity is zero on the rigid sphere:

\begin{figure*}[t]
\begin{center}
\includegraphics[width=16cm]{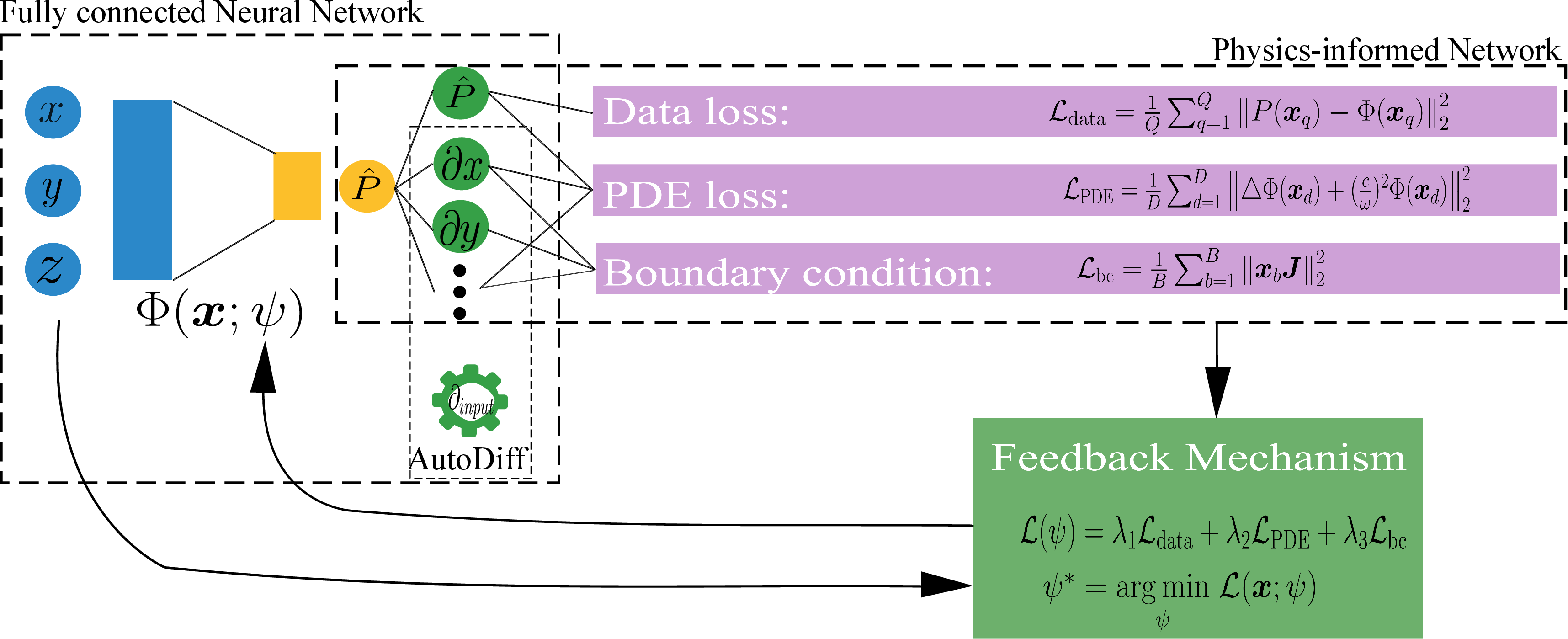}
\end{center}
\caption{The workflow includes three components: an FNN for mapping inputs to outputs with trainable parameters $\psi$, a Physics-Informed Network that incorporates physical governing equations and boundary conditions, and a feedback mechanism to minimize the loss function and update $\psi$.}
\end{figure*}

\begin{equation}
\left.\frac{\partial P(\boldsymbol{r},\omega)}{\partial r}\right|_{r=a}=0.
\label{eq:9}
\end{equation}
It represents the derivative of the sound pressure with respect to the radius of the sphere and can be written in terms of derivatives with respect to the Cartesian coordinates as
\begin{equation}
\frac{\partial P(\mathbf{r}, \omega)}{\partial r} = \frac{\partial P(\mathbf{r}, \omega)}{\partial x} \frac{\partial x}{\partial r} + \frac{\partial P(\mathbf{r}, \omega)}{\partial y} \frac{\partial y}{\partial r} + \frac{\partial P(\mathbf{r}, \omega)}{\partial z} \frac{\partial z}{\partial r}.
\label{eq:10}
\end{equation}
Conduct the substitutions
$\frac{\partial x}{\partial r} = \frac{x}{r}$,
$\frac{\partial y}{\partial r} = \frac{y}{r}$,
$\frac{\partial z}{\partial r} = \frac{z}{r}$,
\eqref{eq:9} can be expressed in a matrix form 
\begin{equation}
\left.\frac{\partial P(\boldsymbol{r},\omega)}{\partial r}\right|_{r=a}
=
\begin{bmatrix}
x& y & z
\end{bmatrix}
\begin{bmatrix}
\frac{\partial P(\mathbf{r}, \omega)}{\partial x} \\
\frac{\partial P(\mathbf{r}, \omega)}{\partial y} \\
\frac{\partial P(\mathbf{r}, \omega)}{\partial z}
\end{bmatrix}=0.
\label{eq:11}
\end{equation}
We denote $\boldsymbol{x_b}$ as a point on the rigid sphere with radius $a$ and $\begin{bmatrix}
\frac{\Phi(\boldsymbol{x};  \psi)}{\partial x} \
\frac{\Phi(\boldsymbol{x};  \psi)}{\partial y} \
\frac{\Phi(\boldsymbol{x};  \psi)}{\partial z}
\end{bmatrix}^{\intercal}$ as $\boldsymbol{J}$. Therefore, we require the PINN estimation to satisfy \eqref{eq:11}, i.e.,
\begin{equation}
\boldsymbol{x_b } \boldsymbol{J}=0.
\label{eq:12}
\end{equation}

At present, we have constructed an FNN and acquired physical knowledge, which allows us to develop a PINN. 
Leveraging automatic differentiation within the neural network, we can efficiently compute the first-order and second-order derivatives of $\Phi(\boldsymbol{x};  \psi)$ as required in (\ref{eq:8}) and (\ref{eq:12}). 
These derivatives are essential for updating the network parameters $ \psi$ using gradient descent.

We formulate the loss function to include three terms, each contributing to the overall training process
\begin{equation}
\begin{split}
\mathcal{L}( \psi) &=\lambda_{1}\underbrace{\frac{1}{Q} \sum_{q=1}^Q\left\|P(\boldsymbol{x}_q)-\Phi(\boldsymbol{x}_q;\psi)\right\|_2^2}_{\mathcal{L}_{\mathrm{data}}} \\
&+\lambda_{2}\underbrace{\frac{1}{D} \sum_{d=1}^D\left\| \triangle \Phi(\boldsymbol{x}_d;\psi)+(\frac{c}{\omega})^2\Phi(\boldsymbol{x}_d;\psi)\right\|_2^2}_{\mathcal{L}_{\mathrm{PDE}}}\\
&+\lambda_{3}\underbrace{\frac{1}{B} \sum_{b=1}^B\left\| \boldsymbol{x}_b
\boldsymbol{J}\right\|_2^2}_{\mathcal{L}_{\mathrm{bc}}}   \ ,
\end{split}
\end{equation}
where $\|\cdot\|_2$ is the 2-norm, $\mathcal{L}_{\mathrm{data}}$ represents the discrepancy between the PINN estimations $\Phi(\boldsymbol{x})$ and the actual measurements $P(\boldsymbol{x}_q)$ obtained from microphones. $\mathcal{L}_{\mathrm{PDE}}$ ensures that the Helmholtz equation is satisfied for the positions $\boldsymbol{x}_d$ within the domain $\mathcal{V}$, and $\mathcal{L}_{\mathrm{bc}}$ imposes the boundary condition of zero radial velocity on the rigid sphere. $D$ are positions surrounding the sphere, and $B$ are positions on the rigid sphere. 
To control the relative importance of these terms and ensure similar influences on the training process, we introduce weight factors $\lambda_{1}$, $\lambda_{2}$, and $\lambda_{3}$.

The weights in the loss function significantly influence the convergence and accuracy of the PINN. By balancing loss weights, a compromise can be achieved between accurately fitting the microphone measurements and adhering to the physical constraints.
Since all three losses aim to approach zero, direct normalization is not feasible.  $\mathcal{L}_{\mathrm{data}}$ is influenced by data noise. To address the issue, we balance $\mathcal{L}_{\text{PDE}}$ and $\mathcal{L}_{\text{bc}}$ to the same order of magnitude during the initial stages of training. Specifically,  We set $\lambda_{1}$ to 1, $\lambda_{2}$ to $(\frac{c}{\omega})^2$, and $\lambda_{3}$ to $r$. 

\begin{figure*}[t]
  \centering
  \begin{subfigure}[b]{0.245\linewidth}
    \includegraphics[width=1.1\linewidth]{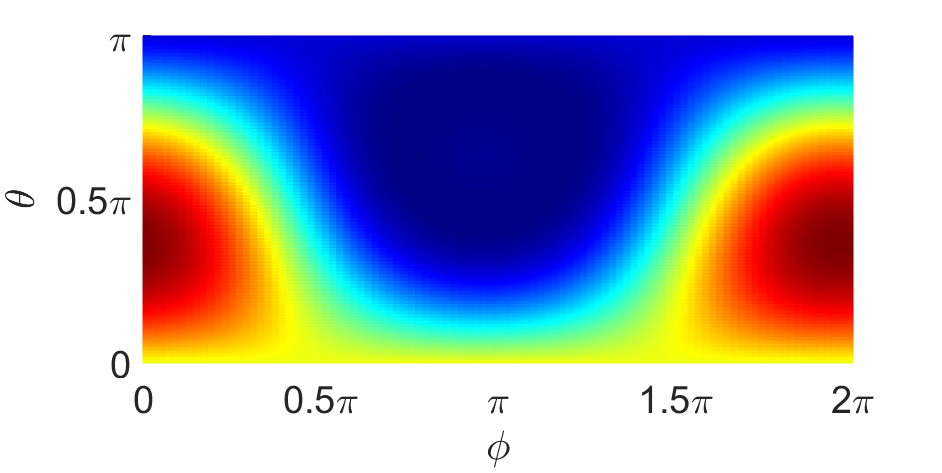}
    \caption{ground truth}
  \end{subfigure}
  \hfill
  \begin{subfigure}[b]{0.245\linewidth}
    \includegraphics[width=1.1\linewidth]{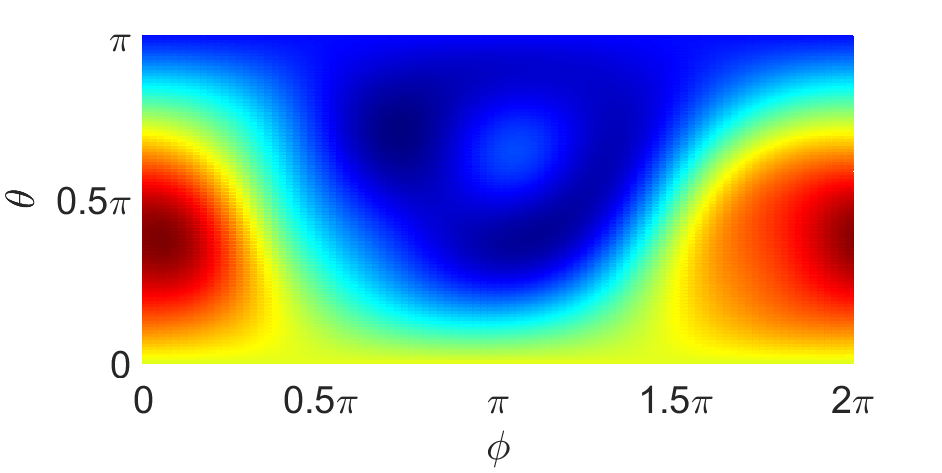}
    \caption{SH estimation}
  \end{subfigure}
  \hfill
  \begin{subfigure}[b]{0.245\linewidth}
    \includegraphics[width=1.1\linewidth]{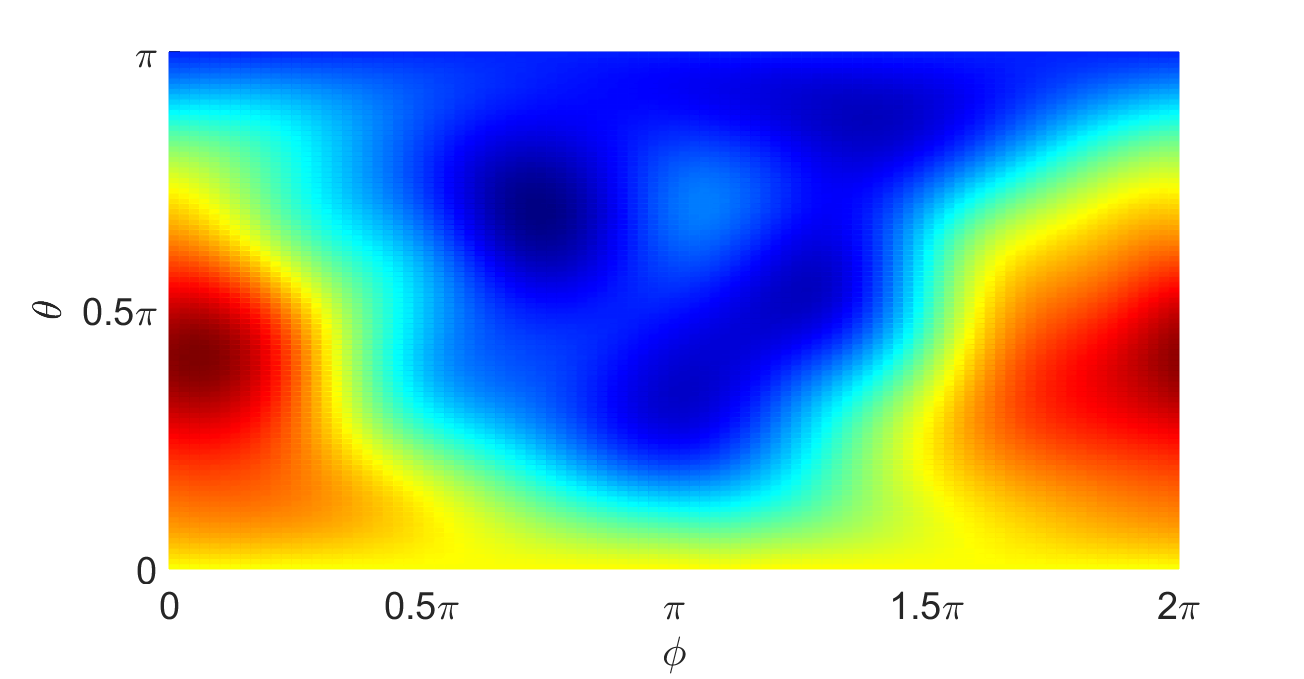}
    \caption{PL estimation}
  \end{subfigure}
  \hfill
  \begin{subfigure}[b]{0.245\linewidth}
    \includegraphics[width=1.1\linewidth]{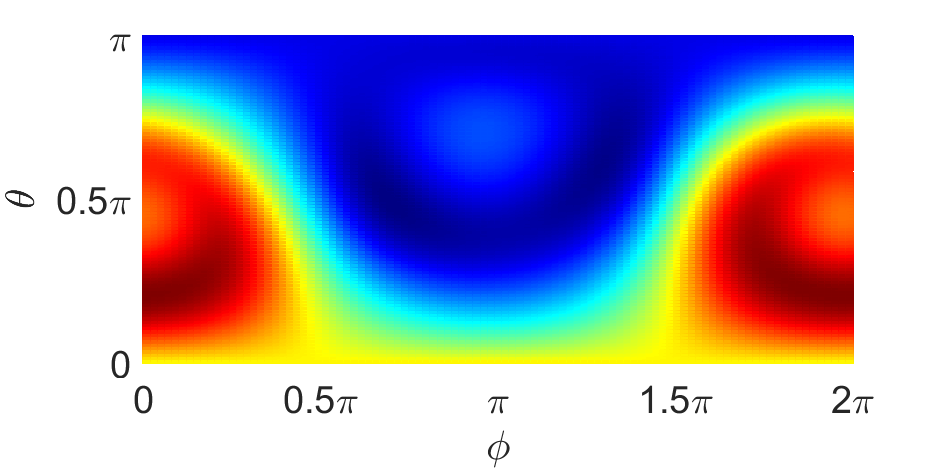}
    \caption{PINN estimation}
  \end{subfigure}
  
  \begin{subfigure}[b]{0.245\linewidth}
    \includegraphics[width=1\linewidth]{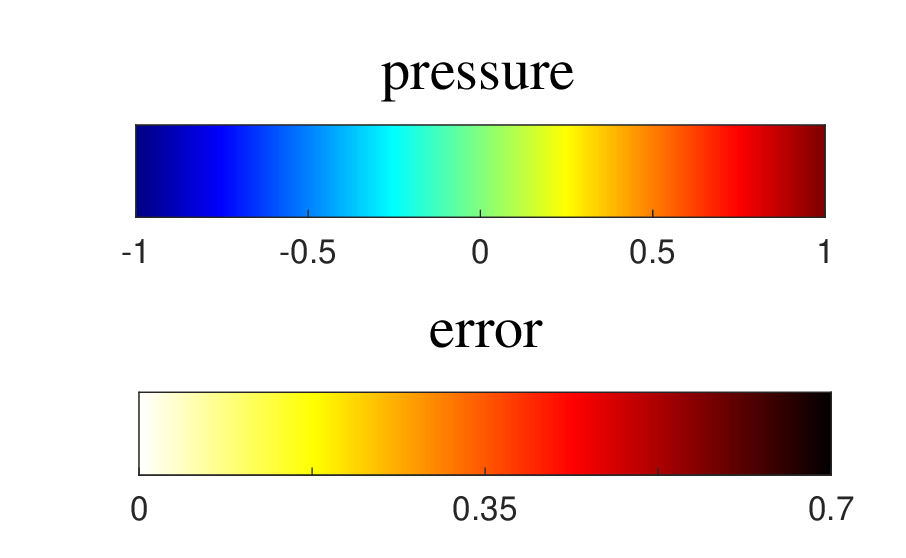}
    \caption{}
  \end{subfigure}
  \hfill
  \begin{subfigure}[b]{0.245\linewidth}
    \includegraphics[width=1.1\linewidth]{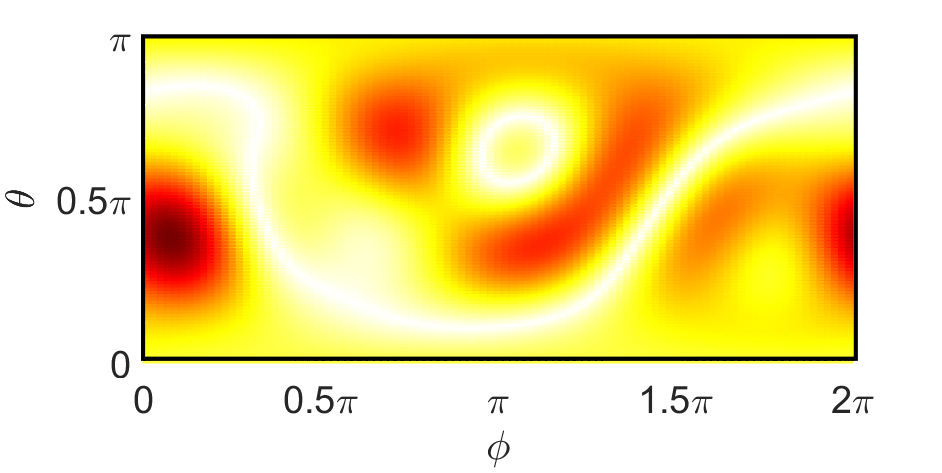}
    \caption{SH error}
  \end{subfigure}
  \hfill
  \begin{subfigure}  {0.245\linewidth}
    \includegraphics[width=1.1\linewidth]{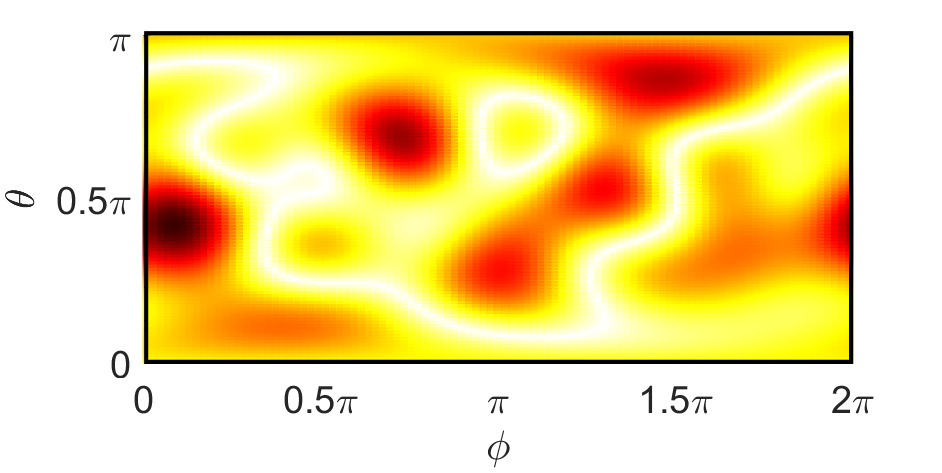}
    \caption{PL error}
  \end{subfigure}
  \hfill
  \begin{subfigure}[b]{0.245\linewidth}
    \includegraphics[width=1.1\linewidth]{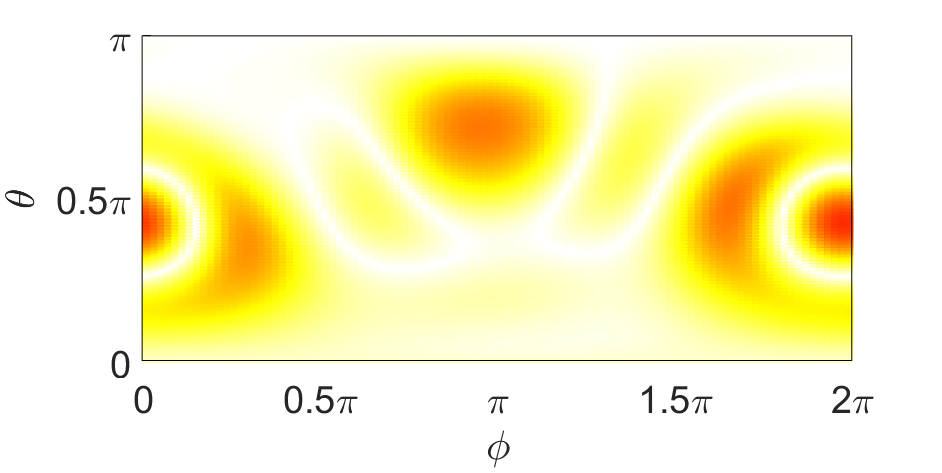}
    \caption{PINN error}
  \end{subfigure}
  \caption{Sound field at $f = 1000$ Hz, $r=0.072$ m (a) The ground truth, (b) SH estimation, (c) PL estimation, and (d) PINN estimation. (f), (g), (h) represent the corresponding errors, respectively, and (e) shows the color bar for reference.}
\label{figgt}
\end{figure*}

\begin{figure}[t]
\begin{center}
\includegraphics[width=9.5cm]{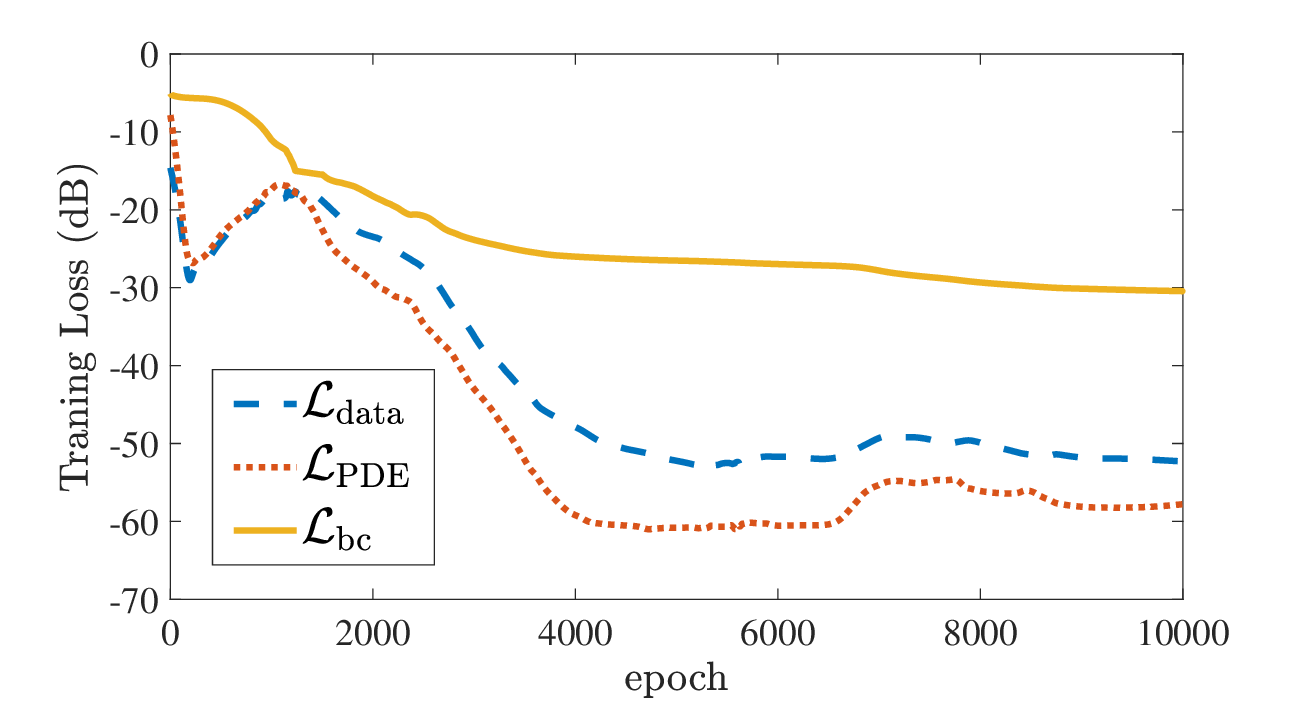}
\end{center}
\caption{PINN training loss: data loss, PDE loss, and boundary condition loss.}
\label{figloss}
\end{figure}

\begin{figure}[t]
\includegraphics[width=9cm]{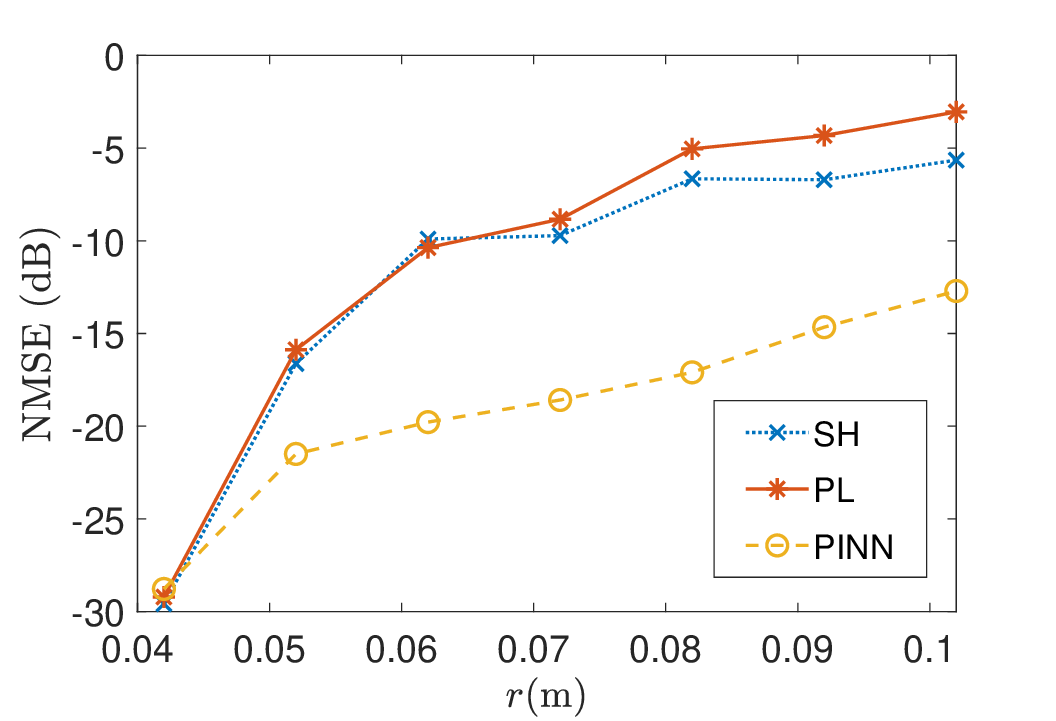}
\caption{The sound field estimation error NMSE at $1000$ Hz as a function of radius.}
\end{figure}

\section{\label{sec:2} simulation}
We conduct simulations in a 3D free field to evaluate the proposed method regarding the sound field estimation surrounding a rigid sphere and compare the performance of three methods: the spherical harmonic method (SH) \cite{abhayapala2002theory}, the plane wave decomposition method (PL) \cite{park2005sound}, and the proposed method (PINN). 

We uniformly mount 32 omnidirectional microphones on a rigid sphere with a radius of $a=0.042$ m as shown in Fig.~1.
We simulate the transfer functions between the sound source and the microphones on the 
rigid sphere according to \cite{rafaely2015fundamentals}
Specifically, we place two point sources at coordinates $(2.5, 0.8, 0.0) \mathrm{~m}$ and $(-2.0, -0.6, 1.2) \mathrm{~m}$, both with an amplitude of 1. The speed of sound is set to $c=343 \mathrm{~m/s}$.
In order to account for realistic conditions, we add white Gaussian noise to the observation signals, resulting in a signal-to-noise ratio (SNR) of $30 \mathrm{~dB}$. The pressure values are then normalized within the range of $[-1, 1]$. 

For the PINN method, we construct a 3-layer FNN with 4 nodes with the hyperbolic tangent (tanh) activation function. To train the PINN, we employ the Adam optimizer \cite{kingma2014adam} with a learning rate of $10^{-5}$ for a total of 10,000 epochs. We set $\mathcal{L}_{\mathrm{data}}$ based on the 32 microphone measurements, $\mathcal{L}_{\mathrm{PDE}}$ based on 1000 randomly distributed points surrounding the sphere, and $\mathcal{L}_{\mathrm{bc}}$ based on 500 uniformly distributed sampling points on the rigid sphere. As we train the network in Cartesian coordinates, we calculate the spherical coordinates $\boldsymbol{r}$ and then convert them to Cartesian coordinates $\boldsymbol{x}$.

In Fig.~\ref{figloss}, we illustrate the training process. It demonstrates the simultaneous and nearly synchronized reduction of the training error $\mathcal{L}_{\mathrm{PDE}}$ and $\mathcal{L}_{\mathrm{bc}}$ in the initial stages of training for the first 2000 epochs. Subsequently, a gradual decline in $\mathcal{L}_{\mathrm{data}}$ is observed. This observation suggests that the weight configurations of $\lambda_{1}=1$, $\lambda_{2}=(\frac{c}{\omega})^2$, and $\lambda_{3}=r$ achieve a good balance.

We define the pressure error between the true and estimated sound fields as
\begin{equation}
\text {error}:= \|P(\boldsymbol{x}_g, \omega)-\hat{P}(\boldsymbol{x}_g, \omega)\|,
\end{equation}
where $P(\boldsymbol{x}_g, \omega)$ and $\hat{P}(\boldsymbol{x}_g, \omega)$ represent the true and estimated pressures, respectively. 

Fig.~\ref{figgt} presents the ground truth, estimated sound pressure using these three methods, and error at a frequency of $1000$ Hz and a radius of $0.072$ m. It is observed that the SH method exhibits higher errors around $\theta=0.5\pi$, the PL method performs the worst among the three due to its assumption of far-field conditions, which ignore the scattering information. the PINN method achieves the best results overall, although it still experiences relatively higher errors at $\theta=0.5\pi$.

To evaluate the overall performance, we defined the normalized mean squared error (NMSE) as 
\begin{equation}
\text {NMSE}:=10 \log _{10} \frac{\sum_{g=1}^{10000}\|P(\boldsymbol{x}_g, \omega)-\hat{P}(\boldsymbol{x}_g, \omega)\|_2^2}{\sum_{g=1}^{10000}\|P(\boldsymbol{x}_g, \omega)\|_2^2},
\label{eq:mse}
\end{equation}
The estimation was performed at 10,000 evaluation positions uniformly sampled within a sphere with a specific radius and frequency. 

Fig. 5 illustrates the relationship between the radius $r$ of the estimation sphere and the NMSE at a frequency of $1000 \mathrm{~Hz}$.
When $r=a$, which corresponds to estimating the sound pressure on the rigid sphere, all three methods yield comparable results. However, as the radius increases, both the SH method and PL method show a significant surge in error. In contrast, the PINN method exhibits a more gradual increase in error as the radius increases. It has a lower NMSE of $7-10$ dB for $r>0.05$ m than SH and PL methods. This indicates that the PINN method performs better compared to the conventional methods.

\section{Conclusion}
In this study, we propose a PINN method to estimate the sound field around a rigid sphere 
based on the measurement of a small number of microphones mounted on the rigid sphere. 
Our method employs an FNN to model the mapping between space positions and sound pressure. 
We incorporate the Helmholtz equation and the fact that sound pressure cannot generate radial motion at the sphere boundary as part of the loss function. 
This integration of physical knowledge into the network architecture and training process enhances the accuracy of estimation.
A major advantage of our approach is that it relies solely on measured data, 
eliminating the need for grid setup and simulated datasets. 
Simulation results demonstrate that our method outperforms the spherical harmonic method and the plane-wave decomposition method.

It is hard to derive theoretical solutions for the scattering sound of complex geometries, but PINN can use physical and geometric constraints to fit the sound pressure distribution. It allows our method to potentially estimate the sound field of arbitrary geometry.

\bibliography{mybib}
\bibliographystyle{IEEEtran}
\end{document}